\documentclass[12pt]{article}

\usepackage{gensymb}
\usepackage{url}
\usepackage{amssymb,amsmath, amsfonts}
\usepackage{cases}
\usepackage{latexsym}
\usepackage{graphicx}
\usepackage{relsize}
\usepackage{natbib}
\usepackage{rotating}
\usepackage[includeheadfoot,left=1.in,right=1.in,top=1.in,bottom=1.in,bindingoffset=0.in,nohead]{geometry}
\usepackage[onehalfspacing]{setspace}
\usepackage{booktabs}
\usepackage{appendix}
\usepackage[pdfborder={0 0 0}]{hyperref}
\usepackage{float}
\usepackage{xcolor}
\usepackage{caption}
\usepackage{subcaption}
\usepackage{minibox}
\usepackage{floatpag}

\begin{document}
\mathchardef\mhyphen="2D 
	
	\title{Probability Assessments of an Ice-Free Arctic:\\
	Comparing Statistical and Climate Model Projections}

	\author{Francis X. Diebold\\University of Pennsylvania \and Glenn D. Rudebusch \\Federal Reserve Bank of  San Francisco}
	
	\maketitle
	
	\begin{spacing}{1}
		
		\noindent \textbf{Abstract}: The downward trend in the amount of Arctic sea ice has a wide range of environmental and economic consequences including important effects on the pace and intensity of global climate change. Based on several decades of satellite data, we provide statistical forecasts of Arctic sea ice extent during the rest of this century. The best fitting statistical model indicates that overall sea ice coverage is declining at an increasing rate. By contrast, average projections from the CMIP5 global climate models foresee a gradual slowing of Arctic sea ice loss even in scenarios with high carbon emissions. Our long-range statistical projections also deliver probability assessments of the timing of an ice-free Arctic. These results indicate almost a 60 percent chance of an effectively ice-free Arctic Ocean sometime during the 2030s -- much earlier than the average projection from the global climate models.
		\thispagestyle{empty}
				
		\bigskip

		\noindent {\bf Acknowledgments}:   For comments and/or assistance we thank the Co-Editor, two anonymous referees,  Andrew Barrett, Mikkel Bennedsen, Philippe Goulet Coulombe, Rob Engle, Max G\"obel, Camille Hankel, Zeke Hausfather, Andrew Harvey, David Hendry, Eric Hillebrand, Luke Jackson, Robert Kaufmann, Siem Jan Koopman, Akshay Malhotra, Tom Maycock, Zack Miller, Dirk Notz, Claire Parkinson,  Felix Pretis, Gladys Teng, Mike Tubbs, Boyuan Zhang, and the Penn Climate Econometrics Research Group.  We are also grateful to  participants at the 2019 meeting of the American Geophysical Union, the 2020 meeting of the European Geophysical Union, the Oxford University Climate Econometrics Seminar, and the Wharton Energy Economics and Finance Seminar.  The views expressed here are those of the authors and do not necessarily represent the views of others in the Federal Reserve System.
		\bigskip
		
		
		\bigskip
		
		\noindent {\bf Contact}:  fdiebold@sas.upenn.edu, glenn.rudebusch@sf.frb.org
			
	\end{spacing}

\section{Introduction}

The Arctic is warming at least twice as fast as the global average \citep{Osborne2018}. This phenomenon of Arctic amplification in surface air temperature is closely connected to a dramatic multi-decade reduction in Northern sea ice. Indeed, since accurate satellite measurements began in 1978, the extent of Arctic summer sea ice has shrunk by about 40 percent, a loss in area comparable to the western continental United States. This drop in sea ice is one of the most conspicuous warning signs of ongoing climate change, but the reduction of sea ice will also play an important role in determining the pace of future global climate change and has wide-ranging implications for the polar region and the rest of the world. 

At a regional level, diminishing sea ice alters polar ecosystems and habitats and introduces major economic opportunities and risks. For example, new deposits of natural gas, petroleum, and other natural resources will become accessible for extraction, emission, and possible spillage \citep{Petrick2017}. Also, reduced ice coverage facilitates tourism and the use of Arctic shipping lanes, which are shorter than traditional passages via the Suez or Panama Canals. These new routes reduce sailing times but increase Arctic environmental risks from, for example, discharges, spills, and soot deposits \citep{Bekkers2016}. Finally, melting sea ice will have geopolitical consequences for Arctic sea-lane control \citep{ebinger2009}.

Although these proximal Arctic effects are important, the far-reaching implications of diminished Arctic sea ice for regulating global climate and weather are even more consequential. Less sea ice and more open water diminishes the reflectivity (or albedo) of the Arctic region, so that, over time, a greater share of solar heat is absorbed by the earth, which leads to further Arctic amplification and increased temperatures worldwide \citep{Hudson2011}. These higher Arctic temperatures promote thawing and erosion of the polar permafrost, which can result in the release of large amounts of carbon dioxide and methane and provide a significant impetus to further global warming \citep{Tanskietal2019}. Increasing Arctic temperatures also hasten the melting of the Greenland ice sheet, further pushing up sea levels \citep{Trusel2018}. Finally, a warming Arctic and loss of sea ice cover can alter the global dynamics of ocean and air streams, and this effect already appears to be changing weather patterns at sub-polar latitudes \citep{PetoukhovandSemenov2010} and weakening thermohaline ocean circulation including the current that warms Europe \citep{LiuFedorov2019}.

In brief, the loss of Arctic sea ice is not only a stark signal of a changing climate, but it also plays an integral role in the timing and intensity of further global climate change. Not surprisingly then, the downward trend in Arctic sea ice has been the subject of hundreds of research studies. The forecasting literature alone is voluminous and impressive in both methodology and substance.\footnote{Recent research includes \cite{petty2017}, \cite{Onoetal2018}, \cite{Peng2018}, \cite{SerrezeAndMeier2019}, and \cite{Ionita2019}. Ongoing prediction research forums include the Sea Ice Prediction Network at \url{https://www.arcus.org/sipn} and the Polar Prediction Project at \url{https://www.polarprediction.net}.} Although the importance of accurate polar prediction is hard to overstate, substantial uncertainty still remains about the future evolution of sea ice.  Indeed, obtaining a deeper understanding of Arctic sea ice loss has been called a ``grand challenge of climate science" \citep{Katssovetal2010}.

Much forward-looking sea ice analysis has been based on large-scale climate models, which represent of the fundamental physical, chemical, and biological drivers of the earth's climate.  These models attempt to capture the dynamics of the oceans, atmosphere, and cryosphere at a high frequency and a granular level of geographic and spatial detail. Such structural physical models are invaluable for understanding climate variation, determining event and trend attribution, and assessing alternative scenarios.  However, from a forecasting perspective, climate models have generally underestimated the amount of lost sea ice in recent decades (\citealp{Stroeve2007}; \citealp{Stroeve2012}; \citealp{Jahn2016}; and \citealp{Rosenblum2017}). In addition, long-range sea ice projections can differ widely across climate models \citep{StroeveNotz2015}.

Given the global significance of Arctic conditions and the progress yet to be made on specifying structural global climate models, purely \emph{statistical} projections of sea ice are an obvious complementary approach. As a practical matter -- across many disciplines -- parsimonious statistical representations often produce forecasts that are at least as accurate as detailed structural models. Specifically for forecasting Arctic sea ice, there is already some evidence that small-scale statistical models with no explicitly embedded physical science can have some success (\citealp{Guemasetal2016}; \citealp{Wangetal2016}). Therefore, we provide a statistical analysis of the long-run future evolution of Arctic sea ice. Our work is distinguished by its use of intrinsically stochastic ``unobserved components" models, with detailed attention to trend, seasonality, and serial correlation. Based on several decades of satellite data, we provide statistical forecasts of the future loss of Arctic sea ice.\footnote{Our analysis does not examine prediction skill in real-time repeated forecasting, as in \cite{Ionita2019}, but considers very long-range projections at a point in time, as in \cite{Guemasetal2016} .}  Importantly, these forecasts provide probability assessments for a range of long-run outcomes and quantify both model parameter uncertainty and intrinsic uncertainty. Of particular interest are probability assessments of the timing of an ice-free Arctic, an outcome with vital economic and climate consequences (\citealp{Massonnet2012}; \citealp{Snape2014}; and \citealp{Jahn2016}). For this analysis, we also introduce a novel statistical modeling mechanism -- a shadow ice interpretation -- that allows us to readily account for the zero lower bound on the extent of Arctic sea ice in our model.

Our resulting distributional forecasts suggest an ice-free Arctic summer is more likely than not within two decades -- \emph{much} sooner than the projections from many large-scale climate models. In particular, we contrast our statistical forecasts with projections from the ensemble of model simulations conducted for the fifth Coupled Model Intercomparison Project (CMIP5) -- a highly-regarded central source for international global climate model projections. On average, these climate models envisage ice-free Arctic conditions close to the end of the century (assuming a range of business-as-usual carbon emissions paths). Thus, besides their relevance for environmental and economic planning, our probability assessments may also provide a useful benchmark for assessing or calibrating global climate models going forward.

We proceed as follows.  In section \ref{linmod}, we introduce a linear statistical trend model and use it to produce long-range sea ice point forecasts.  In section  \ref{shadowice}, we introduce the ``shadow ice" concept to account for the zero-ice lower bound.  In section \ref{quadmod}, we generalize to a nonlinear (quadratic) statistical model and to interval forecasts that incorporate several forms of uncertainty.  In section \ref{GCM}, we compare our statistical model forecasts to global climate model forecasts with particular attention to hard versus soft landings at zero ice.   In section \ref{prob}, we make probabilistic assessments of several sea ice scenarios.  We conclude in section \ref{conclsec}.

\section{A Linear Statistical Model and Point Forecasts}  \label{linmod}

Arctic sea ice has been continuously monitored since 1978 using satellite-based passive microwave sensing, which is unaffected by cloud cover or a lack of sunlight. For a polar region divided into a grid of individual cells, the satellite data provide cell-by-cell brightness readings, which can be converted into fractional ice surface coverage estimates for each cell. Sea ice extent, $SIE$ -- a very commonly-used measure of total ice area -- is the total area of all cells with at least 15 percent ice surface coverage. That is, $SIE$ rounds down cells with measured coverage of less than 15 percent to zero and rounds up cells that pass the 15 percent threshold to full coverage. The up-rounding in $SIE$ is effectively a useful bias correction, as melting pools on summer ice surfaces can be mistaken for ice-free open water. For this reason, we follow common practice and use monthly average $SIE$ data from November 1978 through October 2019 from the National Snow and Ice Data Center (NSIDC).\footnote{Another measure of Arctic ice is sea ice area ($SIA$), which adds together the measured fractions of ice-covered areas of all cells that pass the 15 percent threshold.  For brevity, we report results only for $SIE$, but results for $SIA$ are qualitatively identical.} The NSIDC data use the NASA team algorithm to convert the satellite microwave brightness readings into measured ice coverage \citep{Fettereretal2017}.\footnote{We interpolate the missing December 1987 and January 1988 observations with fitted values from a regression on trend and monthly dummies estimated using the full data sample.}$^,$\footnote{For a comparison of algorithms, and evidence of strong performance of the NASA team algorithm, see \cite{IcePlus}.}

\begin{figure}[t]
	\caption{Arctic Sea Ice Extent ($SIE$) and Fitted Linear Trend}
	\begin{center}
		\includegraphics[trim= 6mm 0mm 0mm 0mm, clip, scale=.13]{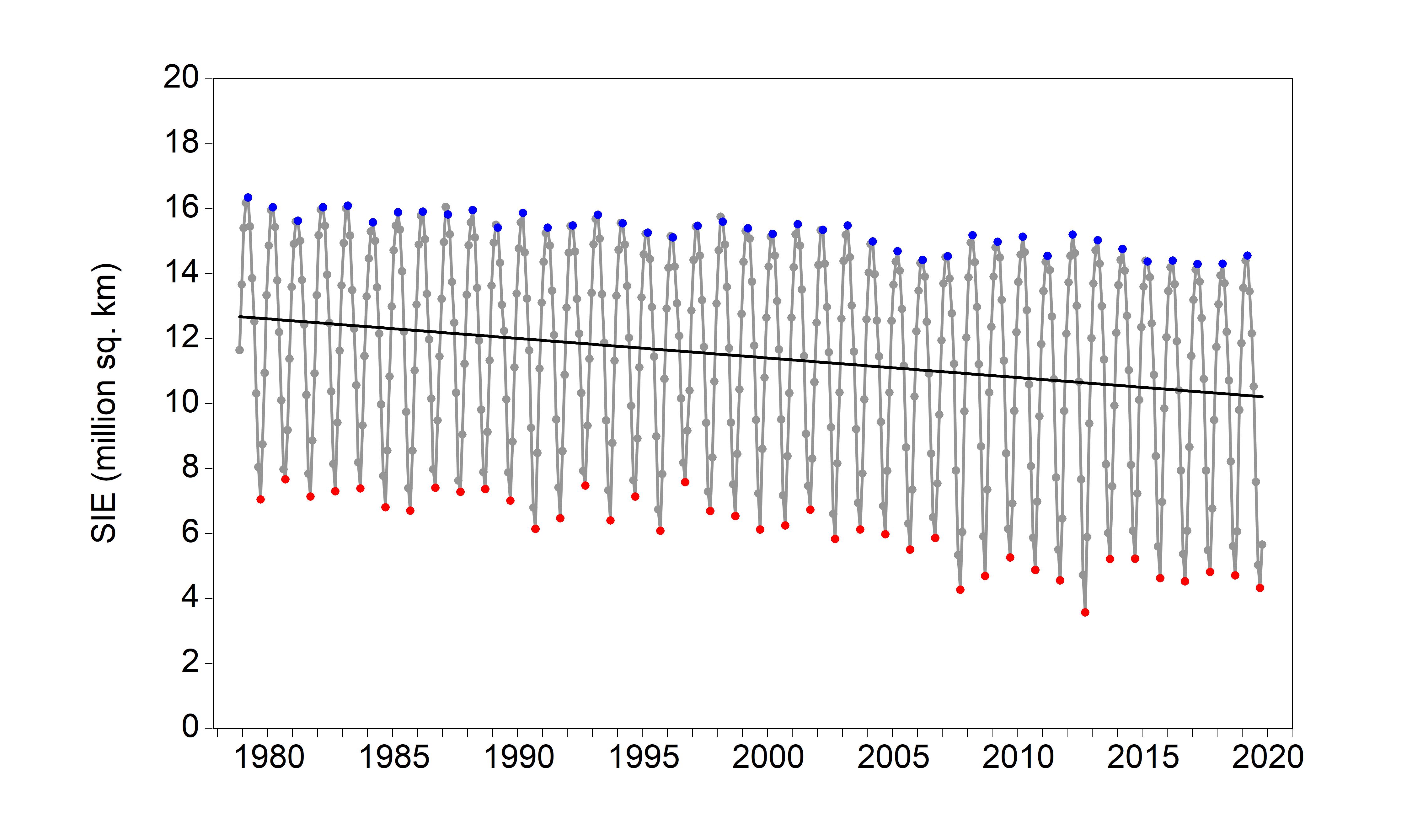}
		\label{tsplot}
	\end{center}
	\begin{spacing}{1.0} \footnotesize \noindent Notes: We show monthly average Arctic sea ice extent ($SIE$) from November 1978 to October 2019 with a fitted linear trend. Each monthly observation is a dot, and September and March observations are colored red and blue, respectively.
	\end{spacing}
\end{figure}

Figure \ref{tsplot} plots the time series of Arctic $SIE$ -- each monthly average observation is a dot -- with an overall estimated linear trend superimposed. The clear downward trend is accompanied by obvious seasonality. A more subtle feature is the possible time variation in the seasonal effects, which may be trending at different rates and possibly nonlinearly.  These effects turn out to be of interest in a complete statistical representation of sea-ice dynamics.

\begin{figure}[t]
	\caption{$SIE$: Linear Model Fits and Point Forecasts}
  		\label{forec_lin}
	\begin{center}
		\includegraphics[scale=0.11]{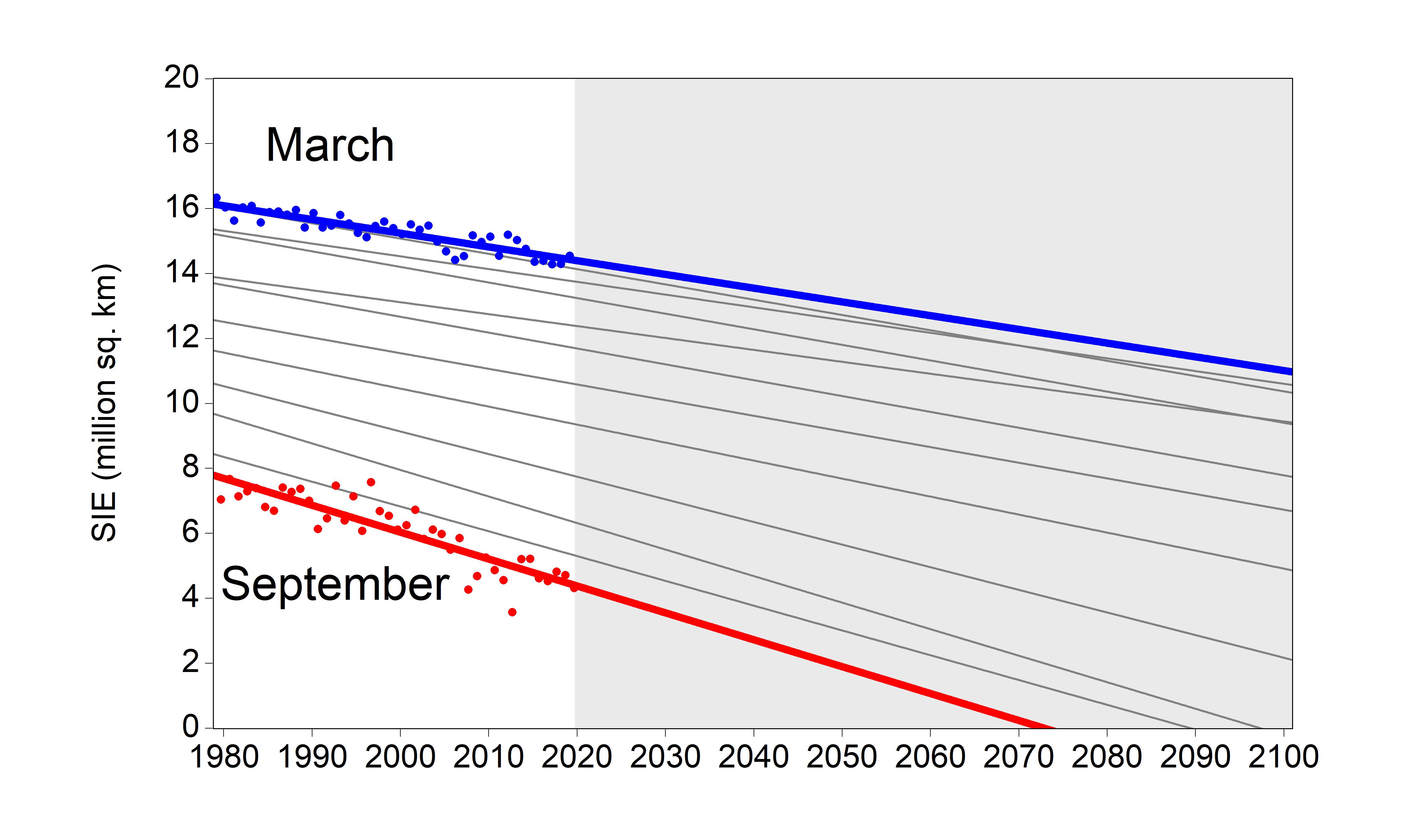}
	\end{center}
	\begin{spacing}{1.0} \footnotesize \noindent Notes: The lines are the twelve in-sample fitted trends and out-of-sample extrapolations for Arctic  $SIE$ in each month based on the linear model (\ref{eq:SIE}). The trends for the months with maximal and minimal sea ice extent are colored -- March in blue and September in red -- and the blue and red dots are the March and September historical $SIE$ data. The out-of-sample period is shaded gray.
	\end{spacing}
\end{figure}

A simple initial representation to capture this variation is a linear statistical model with twelve intercepts, one for each month, each of which may be differently trending, and potentially serially correlated stochastic shocks:
\begin{equation}
	\label{eq:SIE}
SIE_t = \sum_{i=1}^{12} \delta_i D_{it} + \sum_{j=1}^{12} \gamma_j D_{it} {\cdot} TIME_t + \varepsilon_t
\end{equation}
$$
 \varepsilon_t =  \rho \varepsilon_{t-1} + v_t
 $$
 $$
 v_t \sim iid (0, \sigma^2),
 $$
where the $D_{i}$'s are monthly dummy variables ($D_{it}{=}1$ in month $i$ and 0 otherwise, $i{=}1, ..., 12$) and $TIME$ is a time dummy ($TIME_t {=} t$).\footnote{Earlier work in this tradition includes \cite{Peng2018}.}$^,$\footnote{Note that although we allow for seasonal and trending intercepts, we assume a constant disturbance variance across months and time.  There are so many interesting directions that we have reserved volatility modeling for future work.   For example, high-frequency (daily) $SIE$ data are available, from which one could  accurately estimate, and then model, lower-frequency (monthly) realized $SIE$  volatility.}  We estimate model  (\ref{eq:SIE}) -- and other versions below -- by maximizing the Gaussian likelihood.  Detailed regression results for model  (\ref{eq:SIE}) are in column (6) of Table \ref{vvv1} in Appendix \ref{App1}.

Figure \ref{forec_lin} shows the resulting linear trends for all twelve months, highlighting March in blue and September in red.
All of the monthly trends slope downward -- an indication of a warming climate -- and are highly significant. The slopes of the linear trends also differ across months (\citealp{SerrezeAndMeier2019}; \citealp{Cavalierietal2012}).  In particular, the seasonally low-ice months of July through October have  notably steeper downward sloping trends than the seasonally high-ice months of December through May. The estimated September trend, for example, is \textit{twice} as steep as the March trend, and the difference is highly statistically significant.  These linear trends are also extrapolated out of sample (shaded gray) through the end of the century.  For example, September sea ice extent is projected to reach zero just after 2072.

Such linear point forecasts are a useful first step,  but they can be improved by allowing for nonlinearity in the trends and by quantifying forecast uncertainty -- as described in section \ref{quadmod}.  First, however, we elucidate a ``shadow ice" modeling approach that takes into account the fact that the measured amount of sea ice is bounded below by zero.

\section{A Shadow Ice  Interpretation}  \label{shadowice}

One consideration for downward trending statistical models for Arctic sea ice is that the measured amount of sea ice will always be non-negative. By contrast, extrapolations of simple trend models will eventually push into negative territory. There are various functional forms that can be used to model such bounded time series, and the appropriate representation depends very much on the details of the real-world phenomenon under examination.\footnote{One modeling approach is to rescale the bounded time series data to the real line using, say, a log-ratio transformation \citep{Wallis1987}. Alternatively, a time series can be modeled in the original bounded sample space using, for example, the beta autoregressive model of \cite{Rocha2008}. By contrast, some physical science analyses simply terminate model simulations when zero ice conditions are reached, as in \cite{Obryketal2019}.} Some bounds act like reflecting barriers, so the variable of interest spends very little time at the constraint. Other bounds are absorbing states, and once reached, they may be sustained for some time.

With positive amounts of sea ice, fluctuations in \textit{SIE} can serve as a rough approximation for changes in the amount of thermal energy in the Arctic; that is, hotter and colder ocean surface temperatures are reflected in less or more ice, respectively.\footnote{The amount of sea ice depends on a variety of factors including ocean and air temperature, ocean salinity, cloud cover, and wind, current, and wave action.  However, several studies have identified ocean heat content as a major driver of sea ice coverage, notably, \cite{Arthun2012}, \cite{Schlichtholz2011}, and \cite{Selyuzhenok2020}.} However, this connection breaks down when the ice disappears: While \textit{SIE} is fixed at zero, the surface temperature of the Arctic ocean can continue to warm. Furthermore, the warmer the ocean becomes, the less likely there will be a quick return of sea ice, which is indicative of a partially-absorbing state.\footnote{See \cite{StroeveNotz2018} for a description of the phenomenon by which the ocean must release heat back into the atmosphere before sea ice can form again in winter.} 

To account for this effect, the negative values of  sea ice produced by a statistical model can be viewed as a rough expression of ocean temperature. Thus, we redefine the left-hand side variable of the unconstrained model as a \textit{shadow} surface ice extent, $SIE^*$. We view $SIE^*$ as a notional variable that equals measured surface ice when positive, but that may also go negative to represent further increases in ocean thermal energy more broadly. Formally, to translate negative model-based sea ice values into nonnegative sea ice observations, our shadow ice model modifies the unconstrained model \eqref{eq:SIE} to respect the zero lower bound for ice:

\begin{equation} 	\label{eq:Shadow_a}
SIE^*_t = \sum_{i=1}^{12} \delta_i D_{it} + \sum_{j=1}^{12} \gamma_j D_{jt} {\cdot} TIME_t + \varepsilon_t
\end{equation}
$$ 	\label{eq:Shadow_b}
\varepsilon_t =  \rho \varepsilon_{t-1} + v_t
$$
$$  \label{eq:Shadow_c}
v_t \sim iid (0, \sigma^2)
$$
$$  	\label{eq:Shadow_d}
SIE_t = \max(SIE^*_t, 0).
$$
That is, we now interpret our earlier unconstrained linear   model of surface ice as a model of shadow ice, $SIE^*$, so that the  observed extent of sea ice, \textit{SIE}, is the maximum of $SIE^*$ and zero.\footnote{A similar framework has been successfully applied in finance to model nominal interest rates near their zero lower bound \citep{CR2014, BR2016}.}

The shadow ice model respects the nonlinearity of observed ice at the zero lower bound but retains tractability. It also allows us to translate the long-range forecasts from downward trending models like model \eqref{eq:SIE} -- including distributional projections -- into observed data that are always non-negative.  As a matter of physical interpretation, very negative values of shadow ice extent, $SIE^*$, represent environments in which the thermal content of the Arctic ocean is high enough that an immediate return to a positive \textit{SIE} is unlikely. This shadow ice structure provides an intuitive and simple approximation of the thermodynamics of the Arctic ocean transition between sea ice and open water and serves as a useful modeling tool for observed \textit{SIE} dynamics.\footnote{\cite{Wangetal2016} take a different approach by simply constraining the model by the lower bound (for sea ice concentration in their case), so negative predicted values are set to zero. In a dynamic model with lagged sea ice, this procedure will result in a representation of a physical bound that is much closer to a reflecting barrier.}

\section{A Quadratic Statistical Model and Interval Forecasts}  \label{quadmod}

A downward linear trend is a common representation of the secular decline in Arctic sea ice, but linearity is not assured by the physical science. There are a variety of climate feedback mechanisms that could hasten or retard the pace of sea ice loss. The well-known ice albedo effect occurs as sea ice cover is reduced, and the resulting darker ocean surface absorbs more energy, which in turn further reduces sea ice \citep{Stroeve2012}. This feedback effect amplifies sea ice seasonality and may progressively steepen the downward trend in $SIE$ over time \citep{Schroder2014}. The geography of the perimeter of the Arctic Ocean, which is partially constrained by land that blocks expanding winter ice, becomes less relevant as sea ice shrinks, and relaxing this constraint may allow greater seasonal variation and a steeper downward sea ice trend \citep{SerrezeAndMeier2019}.\footnote{Currently, the southern limit of winter ice is bounded by land except in the Bering Sea, Sea of Okhotsk, East Greenland Sea, Barents Sea, and Baffin Bay.} However, there are offsetting negative feedback mechanisms -- associated, for example, with increased cloud cover -- that could slow the rate of sea ice loss over time \citep{IPCC2019}. Indeed, \cite{StroeveNotz2015} argue against trend amplification in favor of trend constancy (linearity) or trend attenuation. Moreover, as described in the next section, long-range $SIE$ projections from large-scale global climate models appear dominated by feedback mechanisms that slow the rate of September sea ice loss over time.\footnote{More extreme forms of nonlinearity -- such as discontinuous breaks, tipping points, and thresholds -- are possible but viewed as less likely \citep{StroeveNotz2015}. \cite {GoldsteinEtAl18} argue that Arctic sea ice is best modeled by step-like shifting means at fitted breakpoints. However, modified statistical information criteria that properly account for the implicit flexibility of such breakpoints -- following \cite{Hall2013} -- do not favor such a shifting mean models relative to a linear trend.}

The lack of a complete understanding of the drivers of Arctic sea ice recommends consideration of a flexible empirical $SIE$ specification, so we generalize from linear to quadratic trends:
\begin{equation}	\label{eq:Shadow2}
SIE^*_t =   \sum_{i=1}^{12} \delta_i D_{it} + \sum_{j=1}^{12} \gamma_j D_{jt} {\cdot} TIME_t + \sum_{k=1}^{12} \alpha_k D_{kt} {\cdot} TIME_t^2  + \varepsilon_t
\end{equation}
$$
\varepsilon_t =  \rho \varepsilon_{t-1} + v_t
$$
$$
v_t \sim iid (0, \sigma^2)
$$
$$
SIE_t = \max(SIE^*_t, 0).
$$
We label model (\ref{eq:Shadow2}) as the ``general" quadratic model as no constraints are imposed on the twelve quadratic ($\alpha_k$) parameters. The linear model (\ref{eq:Shadow_a}) of course emerges under the constraint $\alpha_{1}  {=}  ...  {=}  \alpha_{12} {=} 0$.

\begin{figure}[t]
	\caption{$SIE$: General Quadratic Model Fits and Point Forecasts}
	\begin{center}
		\includegraphics[scale=0.11]{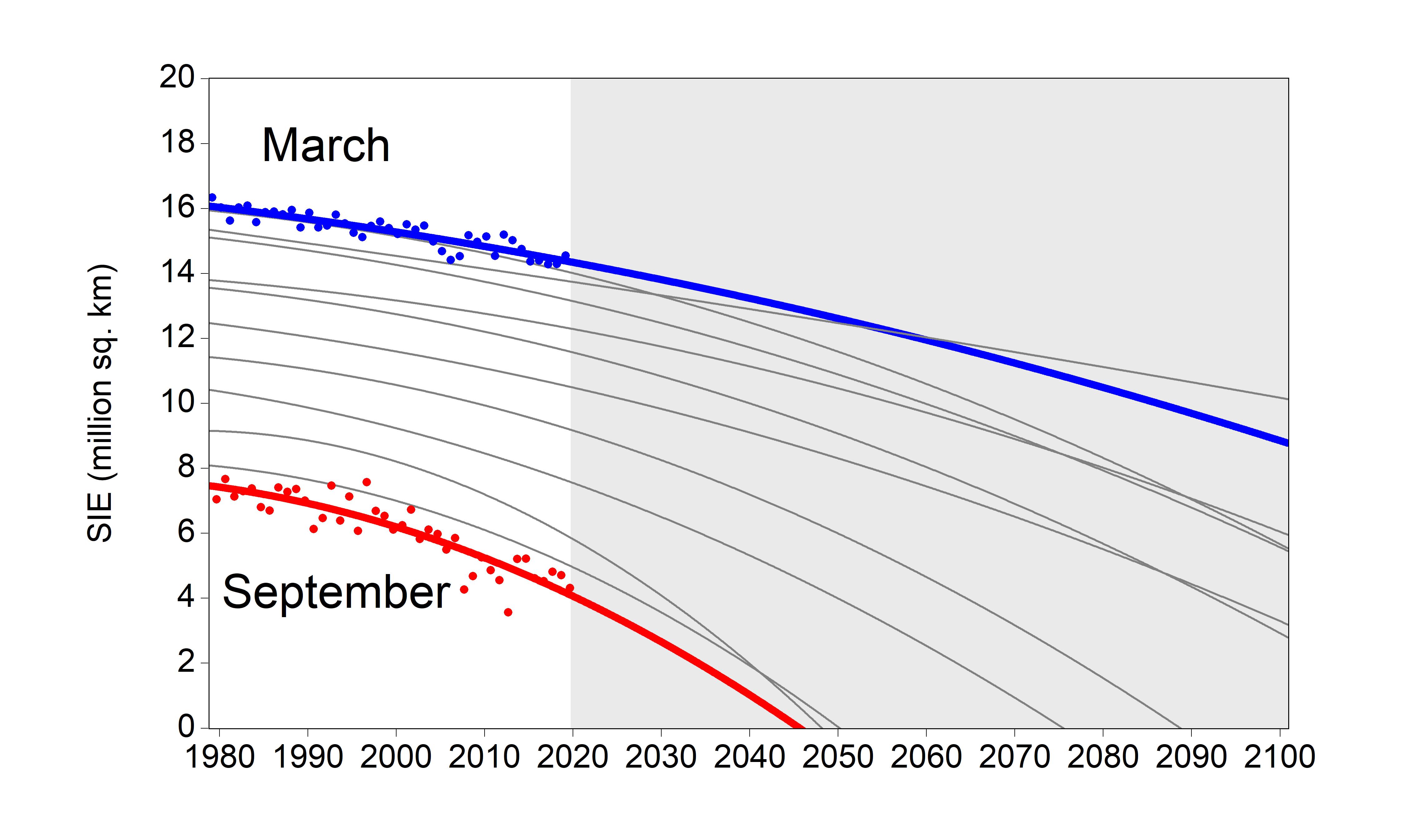}
		\label{forec_quad}
	\end{center}
	\begin{spacing}{1.0} \footnotesize \noindent Notes: The curves are the twelve in-sample fitted trends and out-of-sample forecasts for Arctic  $SIE$ in each month from the general quadratic model (\ref{eq:Shadow2}). The months with maximal and minimal sea ice extent are colored -- March in blue and September in red -- and the blue and red dots are the March and September historical $SIE$ data. The out-of-sample period is shaded gray.
	\end{spacing}
\end{figure}

Figure \ref{forec_quad} shows the $SIE$ estimation fits and forecasts from the general quadratic model, by month, with March and September again highlighted in blue and red, respectively. The trend curvatures for all months are strikingly similar: the trends for all months decrease \textit{at an increasing rate}. That is, the estimated $\alpha_k$ coefficients are negative for every month, indicating that $SIE$ is diminishing at an increasing rate. (Detailed estimation results of model (\ref{eq:Shadow2}) appear in column (1) of Table \ref{vvv1} in Appendix  \ref{App1}.)  The size of the estimated negative coefficients on the quadratic trend terms and their statistical significance vary by month. The most negative and significant $\alpha_k$ coefficients are in the seasonally low-ice ``summer" 
months of August, September, and October, and these months show the greatest trend rates of decline.\footnote{Minimum ice coverage occurs at the end of summer in September because sea ice coverage continues to shrink for any above-freezing temperatures.  For convenience, we refer to the low-ice months of August, September, October as ``summer" even though they span late summer through early autumn.}   An $F$ test of the joint hypothesis that $\alpha_8{=} \alpha_9{=}\alpha_{10}{=}0$ produces a $p$-value of 0.00.\footnote{Because $TIME^2$ and $TIME$  are correlated, an insignificant $\alpha_k$ coefficient would not necessarily imply that nonlinearity is unimportant.} Relative to a linear trend model, the nonlinear trend model forecasts lower sea ice at long horizons.  February-April $SIE$ point forecasts nevertheless remain well above zero through the century, but August-October point forecasts approach zero much more quickly. Indeed the quadratic September point forecast hits zero in 2045.

Table \ref{vvv} summarizes the results from investigation of the summer and non-summer differences in quadratic trend curvature using two standard model selection criteria: the Akaike information criterion (\textit{AIC}) and the Bayesian information criterion (\textit{BIC}). The \textit{AIC} and \textit{BIC} are estimates of out-of-sample forecasting performance (mean-squared error), formed by penalizing estimates of in-sample forecasting performance for degrees of freedom used in model fitting and differing only in the precise penalty applied \citep{Diebold2007}. The table reports these model selection criteria for six versions of the quadratic trend model (\ref{eq:Shadow2}) with various equality constraints imposed on the quadratic coefficients $\alpha_1, ..., \alpha_{12}$. Corresponding estimation results appear in Table \ref{vvv1} in Appendix  \ref{App1}.

The models favored by \textit{AIC} and \textit{BIC} -- that is, those with smaller values -- are very similar and involve summer and non-summer restrictions.  The \textit{AIC} selects a model with equal $\alpha_k$'s for the nine non-summer months (November-July) and unconstrained $\alpha_k$'s for the three summer months (August-October).  The \textit{BIC}, which penalizes degrees of freedom more harshly, selects a slightly more constrained model, with the non-summer $\alpha_k$'s again constrained to be equal and the three summer $\alpha_k$'s \textit{also} constrained to be equal. Compared to the unconstrained version of equation (\ref{eq:Shadow2}) -- the general model -- both the \textit{AIC} and \textit{BIC} prefer specifications with some summer and non-summer equality constraints imposed, because their imposition saves degrees of freedom without substantially degrading fit.  All told, the results of  Table \ref{vvv}  suggest a ``simplified" quadratic model, namely model (\ref{eq:Shadow2}) with both summer and non-summer quadratic coefficients constrained separately to equality: $\alpha_{8} {=} \alpha_{9}  {=}  \alpha_{10}$ and $\alpha_{11}  {=}   \alpha_{12}  {=}  \alpha_{1}  {=}  ...  {=}  \alpha_{7}$.  The regression diagnostics reported in Appendix \ref{App1} suggest good performance of the simplified quadratic model: $R^2$ is almost perfect;  the residual graph and Durbin-Watson statistic (DW) are consistent with random noise; and residual skewness, kurtosis, and density estimates are consistent with normality.  All told, the model appears to do  a fine job of capturing $SIE$ signal, reducing its complicated dynamics to Gaussian white noise.

\begin{table} [t]
		\caption{Akaike and Bayes Information Criteria for Quadratic Coefficient Constraints}
		\label{vvv}
			\begin{center}
			{
			\begin{tabular}{ c  c c  c c  c  c}
			\toprule
			&(1)&(2)&(3)&(4)&(5)&(6)\\
			& NONE & Seq  &   NSeq &  Seq+NSeq & ALLeq & ALL0    \\
			\midrule
			\textit{AIC} &  -0.0673 [3] & -0.0651 [4]  &     \textbf{-0.0913 [1]}     &    \textbf{-0.0877 [2]}     & -0.0639 [5]  & -0.0569 [6] \\
			\textit{BIC} &  \phantom{-}0.2569 [6] &  \phantom{-}0.2421 [5]  &       \textbf{\phantom{-}0.1647 [2]}  &     \textbf{\phantom{-}0.1513 [1]} &  \phantom{-}0.1665 [4] & \phantom{-}0.1649 [3]\\
			\bottomrule	
		\end{tabular}
	}
	\end{center}

	\begin{spacing}{1.0} \footnotesize \noindent Notes:  We show \textit{AIC} and \textit{BIC} values for the quadratic model with various equality constraints imposed  on the quadratic coefficients $\alpha_1, ..., \alpha_{12}$. ``NONE" denotes no constraints, which corresponds to the general quadratic model (\ref{eq:Shadow2}).  ``Seq" (``Summer equal") denotes August-October equal  ($\alpha_{8} {=} \alpha_{9}  {=}  \alpha_{10}$).  ``NSeq" (``Non-Summer equal") denotes November-July equal ($\alpha_{11}  {=}   \alpha_{12}  {=}  \alpha_{1}  {=}  ...  {=}  \alpha_{7}$). ``Seq+NSeq" denotes summer months equal and non-summer months (separately) equal, which corresponds to the simplified quadratic model.  ``ALLeq " denotes all months equal ($\alpha_{1}  {=}  ...  {=}  \alpha_{12}$). ``ALL0" denotes all months 0 ($\alpha_{1}  {=}  ...  {=}  \alpha_{12} {=} 0$),  which corresponds to the linear model (\ref{eq:SIE}).  Model ranks appear in brackets, where [1] denotes the best (smallest) criterion value. We show in boldface the best two models according to each criterion.
\end{spacing}
\end{table}

\begin{figure}[t]
	\caption{$SIE$: Simplified Quadratic Model Fits, Point Forecasts, and Interval Forecasts}
	\begin{center}
		\includegraphics[scale=0.11]{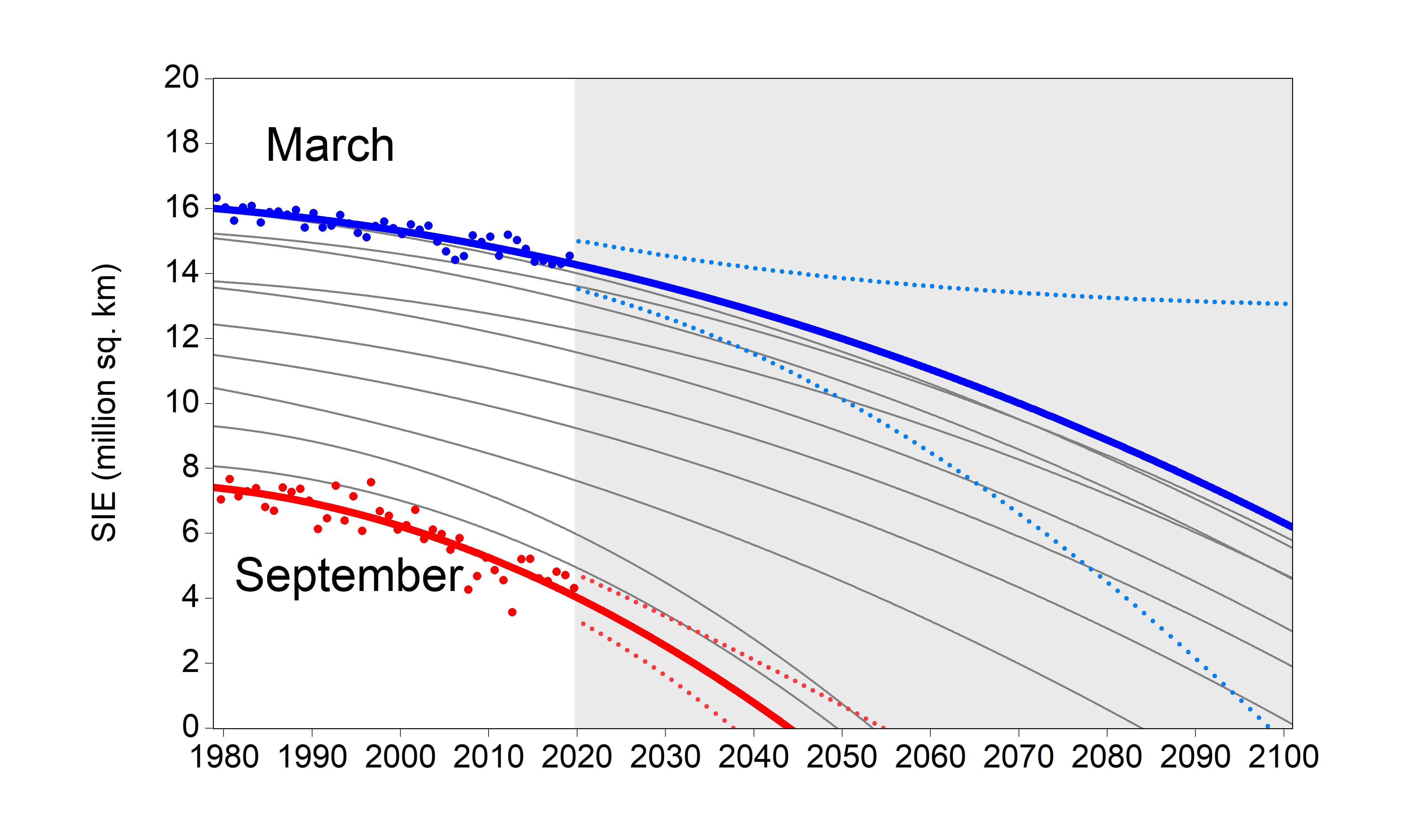}
		\label{forec_quads}
	\end{center}
	\begin{spacing}{1.0} \footnotesize{Notes: The curves are the twelve in-sample fitted trends and out-of-sample forecasts for Arctic  $SIE$ in each month from the simplified quadratic model. The months with maximal and minimal sea ice extent are colored -- March in blue and September in red -- and the blue and red dots are the March and September historical $SIE$ data. The out-of-sample period is shaded gray.  The March and September forecasts also have $\pm \, 2 \, s.e.$ blue and red dotted bands that account for parameter estimation uncertainty and intrinsic uncertainty error.}
	\end{spacing}
\end{figure}

Figure \ref{forec_quads} shows the simplified quadratic model fits and forecasts.  The simplified model point forecasts are very similar to those of the general quadratic model in Figure \ref{forec_quad}, with a zero-ice September also reached in 2045, but the rank ordering of the months in terms of $SIE$ is better preserved going forward. Going beyond these point forecasts, an important advantage of a formal statistical approach is that it can quantify the amount of future uncertainty. Figure \ref{forec_quads} supplements the simplified quadratic trend point forecasts with interval or probability density forecasts. When making long-horizon interval forecasts, it is crucial  to account for parameter  estimation error, because its deleterious effects grow with the forecast horizon.  For example, although parameter estimation error may have small effects on 6-month-ahead intervals, it will be greatly compounded for 600-month-ahead intervals.  An estimate of the time-$t$ standard deviation of the forecast error, which accounts for parameter estimation error, is $\delta_t {=} s \sqrt{1 + x_t' (X'X)^{-1} x_t}$,
where  $s$ is the standard error of the regression, $x_t$ is a 36$\times$1 column vector of time-$t$ right-hand-side variables,  $X$ is a $T$$\times$36 matrix whose columns contain the regression's right-hand-side variables over time, and $T$ is sample size.\footnote{See for example  \cite{Johnston1972}, pp. 153-155, for derivation of this canonical result.}  We use $\delta_t$ to produce the  $\pm \, 2 \, \delta_t$ pointwise prediction intervals of Figure \ref{forec_quads}. Under normality of the shocks underlying the simplified quadratic model, the $\pm \, 2 \, \delta_t$ intervals are approximate 95 percent  confidence intervals.\footnote{Normality of the $v$ shocks does not appear unreasonable, as the simplified quadratic model residuals have skewness and kurtosis of -0.17 and 3.73, respectively. We also obtained similar results without assuming normality via bootstrap simulation, which we discuss in section \ref{prob}.}  The intervals widen rapidly; indeed the September interval starts to include zero before 2040.\footnote{By contrast, interval forecasts that fail to account for parameter estimation uncertainty  quickly approach (by about 12 months ahead) the fixed-width interval  $\pm \, 2 \, \delta$, where $\delta$ is the estimated unconditional standard deviation of the $AR(1)$ disturbance in equation (\ref{eq:Shadow2}), $\sqrt{\hat{\sigma}^2 / (1 {-} \hat{\rho}^2)} \approx 0.32$ and fail to widen with forecast horizon.}

\begin{figure}[t]
	\caption{$SIE^*$: Simplified Quadratic Model Point and Interval Forecasts}
	\begin{center}
		\includegraphics[trim= 10mm 30mm 0mm 30mm, clip, scale=.13]{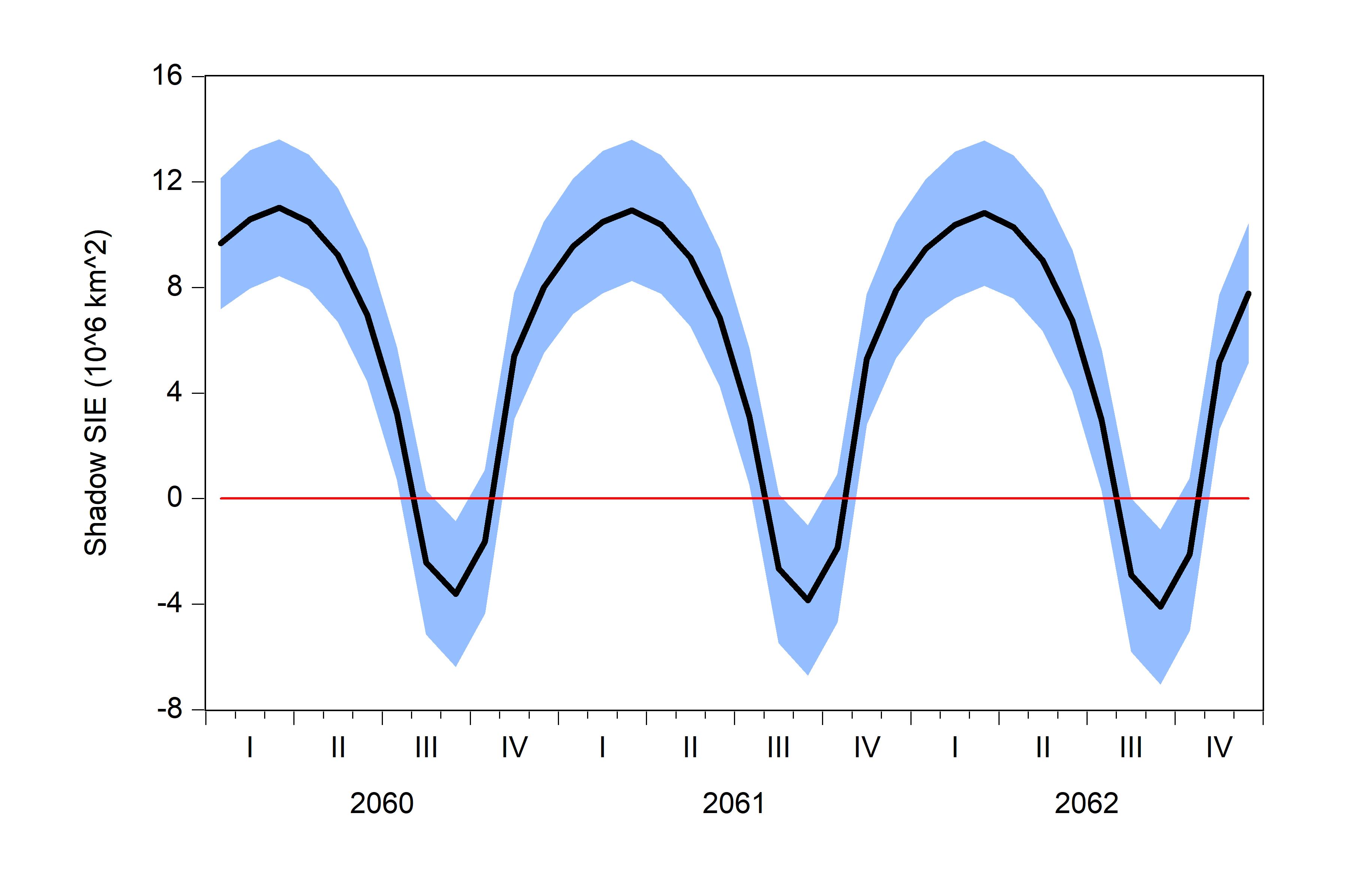}	
		\label{shadow}
	\end{center}
	\begin{spacing}{1.0} \footnotesize Notes: We show the forecast of Arctic shadow sea ice extent, $SIE^*$, for   2060-2062, based on the simplified quadratic model.  The solid black line is the point forecast, the shaded area is the $\pm \, 2 \, s.e.$ band, and the red line denotes zero.
	\end{spacing}
\end{figure}

Finally, in Figure \ref{shadow} we zoom in on the 2060-2062 (36-month) segment of the simplified quadratic model shadow ice forecast.  The point forecast trends down and is seasonally below zero by then. (Indeed, as already discussed,  the point forecast is seasonally below zero well before then.)  Moreover, the $\pm \, 2 \, s.e.$ intervals widen over time, and entire \textit{intervals} are seasonally below zero by then. Hence it appears that, {with near certainty}, summer $SIE$ will vanish by 2060. This result is also clear from our earlier Figure \ref{forec_quads}, but Figure  \ref{shadow}  highlights it in a different and complementary way.

\section{Statistical versus Climate Model Projections}  \label{GCM}

Of the many analyses of the long-term future evolution of Arctic ice, most have focused on  projections from large-scale climate models. Such models are based on the underlying physical, chemical, and biological processes that govern the dynamics of weather and climate across ocean, air, ice, and land. The models fit an immense number of variables at a high temporal frequency and a granular spatial scale (e.g., a 30-minute time interval. a 100km worldwide horizontal resolution, and 40 vertical levels in the oceans and atmosphere). Dozens of scientific groups around the world have constructed and currently maintain such models. Occasionally, these groups conduct concurrent simulations as part of the Coupled Model Intercomparison Project (CMIP) that involve common sets of inputs including carbon emissions scenarios. The most recently completed iteration or phase of this project is the fifth one, denoted CMIP5 \citep{Taylor2012}.  The CMIP5 model comparison study was the main source of climate projections included by the International Panel on Climate Change (IPCC) in its landmark Fifth Assessment Report.

\begin{figure}[tb]
	\caption{September $SIE$: Statistical and Climate Model Fits, Point Forecasts, and Interval Forecasts}
	\begin{center}
		\includegraphics[scale=0.12]{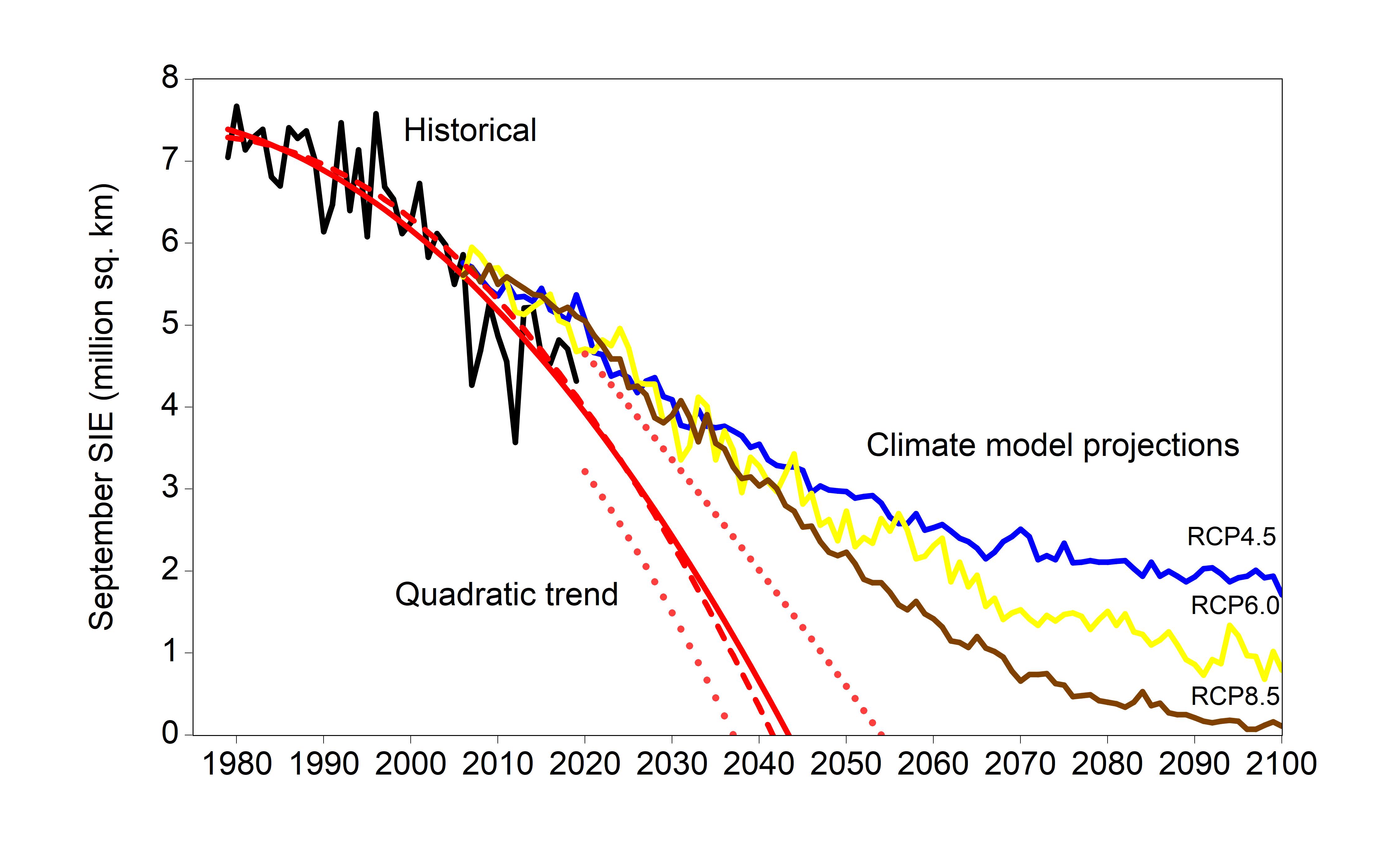}
		\label{CModel}
	\end{center}
	\begin{spacing}{1.0} \footnotesize \noindent Notes: We show fitted values and forecasts of Arctic sea ice extent, $SIE$. The solid black line is the historical data for September Arctic sea ice extent from 1979 to 2019.  Mean projections from CMIP5 climate models are shown assuming underlying pathways of a high level of emissions (RCP8.5, brown), medium emissions (RCP6.0, yellow), and lower emissions (RCP4.5, blue). The climate model projections start in 2006. The red dashed line is the fitted and projected simplified quadratic model estimated using data from 1979 to 2005. The red solid line is the fitted and projected simplified quadratic trend model estimated using the full sample from 1979 to 2019, and the red dotted lines are  $\pm \, 2 \, s.e.$ intervals (approximate 95\% confidence intervals) around that projection accounting for  parameter estimation uncertainty and intrinsic uncertainty. 
		\end{spacing}
\end{figure}

Arctic $SIE$ is a key variable projected by climate models, and the models included in CMIP5 are generally judged to provide a better fit to Arctic sea ice than earlier CMIP iterations.\footnote{\cite{Stroeve2007} describes the poor sea ice fit of the CMIP3 models, while the somewhat improved fit of the CMIP5 models is noted by in \cite{Stroeve2012}.}  Figure \ref{CModel} shows three projections for September $SIE$ constructed as averages across sets of CMIP5 global climate models. These multi-model mean projections are constructed under three different scenarios, or Representative Concentration Pathways (RCPs), for future greenhouse gas concentrations. The brown, yellow, and blue lines are averages of the climate models, respectively, under a high level of carbon emissions (RCP8.5), a medium level (RCP6.0, yellow), and a low level (RCP4.5, blue).\footnote{The climate model data are described in \cite{SerrezeAndMeier2019} and \cite{Stroeve2012} and were kindly provided by Andrew Barrett at the National Snow and Ice Data Center. The model sets averaged for each scenario are not identical, with the RCP4.5, RCP6.0, and RCP8.5 scenarios based on 26, 8, and 25 climate models, respectively.} The first two higher emissions scenarios are viewed as more likely business-as-usual outcomes and are the most relevant to compare to statistical projections that extend the historical sample of past data and assume a continuation of the world economy's current population and development trajectories.\footnote{In the RCP8.5 scenario, continuing increases in greenhouse gas emissions through the end of the century raise the 2100 global average temperature by about 4.0-6.0$^{\circ}$C above pre-industrial levels \citep{USGCRP2018}. In the RCP4.5 scenario, greenhouse gas emissions level off before mid-century, and the 2100 global average temperature is approximately 2.0-3.0$^{\circ}$C above pre-industrial levels. }

The solid red line in Figure \ref{CModel} shows the September trend from the simplified quadratic model estimated on the full sample from November 1978 to October 2019. The dotted lines bracketing this forecast provide an approximate 95 percent confidence interval. However, the CMIP5 climate model projections are only based on data through 2005 and do not include the past dozen or so years of sea ice observations. For comparability to these climate model projections, we re-estimated the simplified quadratic model using data from 1978 through 2005, and the September trend from this pre-2006 model is shown as the dashed red line. An interesting first result is that from the close conjunction of the two simplified model statistical projections (the solid and dashed red lines), the addition of data from 2006 to 2019 does not lead to a significant revision in the statistical trend model. The very modest differences between the pre-2006 estimated quadratic trend and the full-sample version is an indication of the stability and suitability of the simplified statistical model.

Comparing the statistical and climate model projections in Figure \ref{CModel} reveals two salient features. First, throughout the forecast period, the full-sample statistical model projection is significantly lower than any of the climate multi-model mean projections. Indeed, the climate model mean projections are well outside the 95 percent confidence intervals. Relative to the statistical model estimated on data before 2006 or the full-sample version, the climate model means are overestimating the 2019 $SIE$ trend by about 1.0 million km\textsuperscript{2}. This wedge is projected to increase dramatically over the next two decades. Indeed, the pre-2006 and full-sample statistical trend models project zero ice in 2042 and 2044, respectively, but none of the climate model mean projections reach a completely ice-free Arctic in this century. Notably, even assuming a high level of emissions (RCP8.5) -- which is a scenario with continuing increases in average global surface temperatures throughout this century -- the multi-model mean projection never reaches zero Arctic summer sea ice.

Second, in contrast to the statistical projection, the climate model mean projections show a decreasing rate of ice loss over time -- that is, a concave rather than convex structure. Specifically, the climate model projections all display a roughly linear decline for the first couple of forecast decades (through around 2040) and then start to level out. For the lower emissions RCP4.5 scenario, the leveling out and deceleration of sea ice partly reflects a slowdown in the pace of global temperature increases. This effect is not at work in the RCP8.5 scenario, which has global temperatures steadily climbing through 2100. However, very close to zero sea ice extent, the leveling out of $SIE$ in RCP8.5 appears to reflect the hypothesized difficulty of melting the thick sea ice clinging near northern coastlines -- notably in Greenland and Canada. Climate models generally assume that these coastal regions will retain landfast sea ice for a time even after the open Arctic Sea is free of ice. Therefore, a common definition of ``ice-free" or ``nearly ice-free" in the literature is a threshold of 1.0 million km\textsuperscript{2} rather than zero SIE \citep{WangOverland2009}. Still, even with this higher threshold, the three climate model mean projections only reach a nearly ice-free Arctic in 2068, 2089, and $>$2100 for successively lower emissions scenarios, respectively. By contrast, the pre-2006 statistical projection reaches the higher 1.0 million km\textsuperscript{2} threshold in 2037, and the full-sample statistical model reaches that level in 2039.

Some have argued that climate models generally do well in representing the large-scale evolution of Arctic sea ice \citep{StroeveNotz2015}, but a number of studies have noted that the CMIP5 global climate models overpredicted the amount of Arctic sea ice (\citealp{Massonnet2012}; \citealp{StroeveEtAl2012b}; \citealp{SerrezeAndMeier2019}). That overprediction continues, and its source is not well understood. One proposed correction to this overprediction has been to focus on the models that fit the historical $SIE$ data better according to certain metrics \citep{WangOverland2012}. However, there is no agreed upon model selection criterion.  Also, \cite{Rosenblum2017} discount such model selection because models with more accurate sea ice readings also tend to overpredict global temperatures, so the selected climate models may be getting sea ice loss right for the wrong reason. Finally, it should be noted that from a statistical viewpoint, focusing on a simple average forecast from many models has been shown to be a robust prediction strategy (\citealp{DieboldAndShin2019}).

In some respects, the wide differences between the statistical and climate model projections are not too surprising. In climate models, the monthly observations on total Arctic SIE are a high-level output from complex, nonlinear, granular representations of the relevant underlying science.  Obtaining good SIE predictions from these models requires correctly specifying a host of detailed subsidiary processes.  Such a bottom-up modeling procedure has important advantages in structural interpretation and counterfactual scenario analysis.  However, in a variety of disciplines, a bottom-up procedure, which carries the possibility that small misspecifications can accumulate and affect high-level aggregates, has not been found to improve prediction relative to a top-line procedure that directly models the object of interest (\citealp{Diebold2007}).  Thus, based on broad previous experience, we believe that direct statistical projections of Arctic SIE are likely to be relatively accurate.

\section{Probability Assessments of an Ice-Free Arctic}  \label{prob}

An advantage of a formal statistical model is its ability to make probability density forecasts for a range of possible reduced Arctic sea ice scenarios. Of particular interest are probability assessments of an ice-free or nearly ice-free Arctic, that is, the probability that $SIE_t$ equals zero or is less than or equal to some threshold $\gamma$, respectively. Formally, such event probabilities can be denoted as $P(SIE_t{\le} \gamma)$, which represents the probability that sea ice extent is less than or equal to $\gamma$ in month $t$.
We  estimate these scenario probability distributions using the simplified quadratic model and a stochastic simulation procedure that accounts for parameter estimation uncertainty and allows for potentially non-Gaussian serially correlated stochastic shocks. From a given set of simulated paths, we estimate the event probabilities of interest as the proportion of simulated paths in which the event occurs out of the total number of paths.\footnote{We provide details on this simulation procedure in Appendix \ref{App2}. For further discussion of the econometrics of threshold event probabilities, see \cite{BR2016}.} 

\begin{figure}[t]
	\caption{Probability Distributions of First Ice-Free September and First Ice-Free Summer }
	\begin{center}
		\includegraphics[scale=0.131]{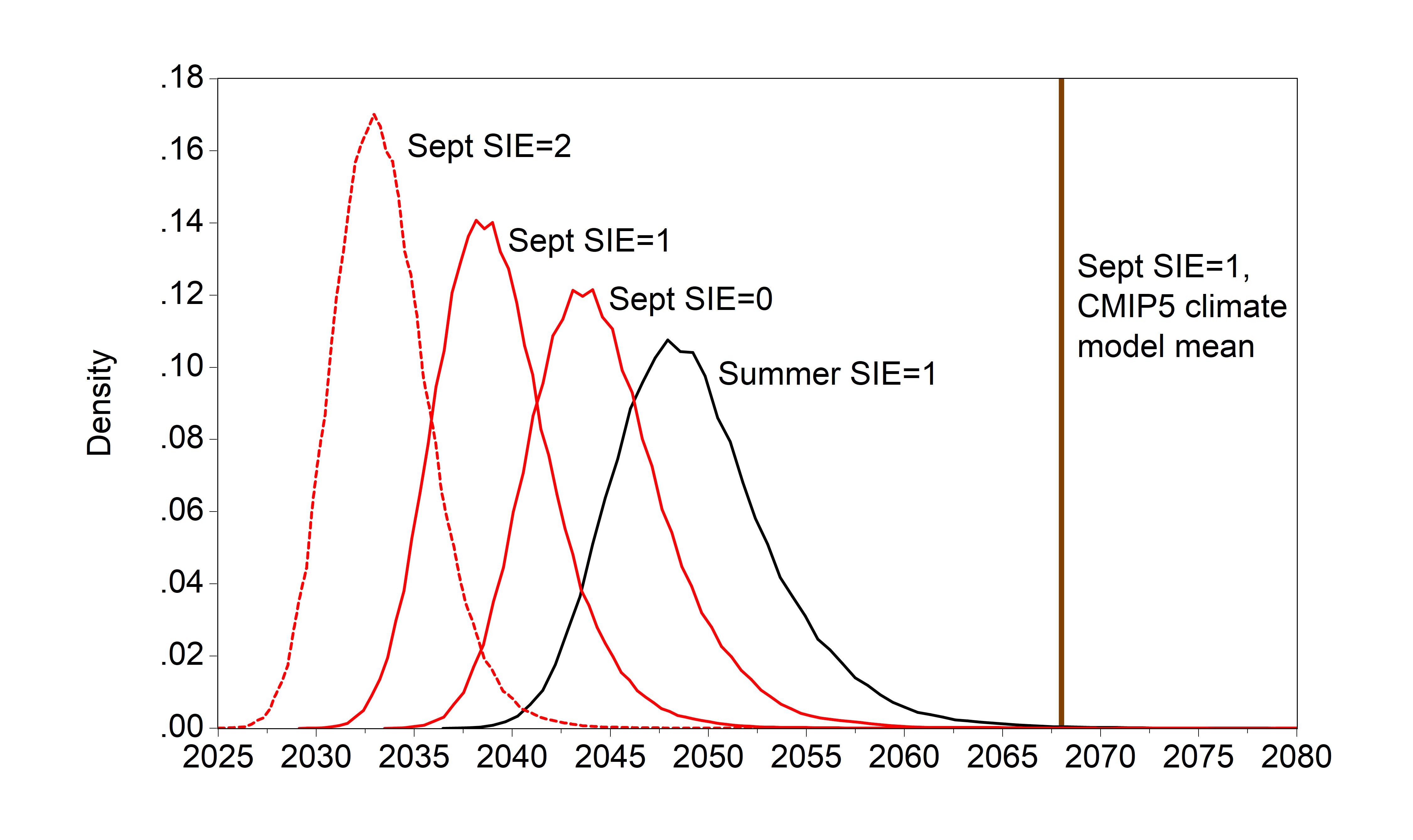}
		\label{densityS}
	\end{center}
	\begin{spacing}{1.0} \footnotesize{Notes: We display probability distributions from the simplified quadratic trend model for the date of the first effectively ice-free September (in red) and first ice-free summer (in black). The ``ice-free" threshold is defined as $SIE = 0$, 1, or 2 million km\textsuperscript{2}. The brown vertical line denotes the date that the mean CMIP5 climate model projection for RCP8.5 reaches $SIE  \le $ 1 million km\textsuperscript{2}.}
	\end{spacing}
\end{figure}

An event that has attracted much attention in the literature is the initial occurrence of an ice-free or nearly ice-free September.  We calculate the probability for each September with date $t_0$ so that $SIE_{t_0}{\le} \gamma$ and $SIE_t {>} \gamma$ for all $t{<}t_0$. Specifically, for a given $\gamma$ and simulation $i$, we determine the year in which September $SIE_t$ first reaches $\gamma$, and then we cumulate across  all simulations to build a distributional estimate.  The red lines in Figure \ref{densityS} provide the resulting probability distributions for an initial ``ice-free" Arctic September for $\gamma{=}0$, $\gamma{=}1$, and $\gamma{=}2$, that is, for progressively more lenient definitions of ``ice-free."   As noted above, the middle value of $\gamma$, which represents Arctic sea ice of less than $1$ million ${\rm km}^2$, is a popular benchmark in the literature \citep{WangOverland2012}. For that particular threshold, the statistical model produces a distribution centered at 2039.\footnote{Both the mean and median round to 2039 despite a slightly skewed distribution toward longer times.} Of particular interest is the probability distribution of dates taking into account model parameter uncertainty and stochastic shock uncertainty, and this distribution shows about a 60 percent probability of an effectively ice-free September Arctic occurring in the 2030s.

The distribution with $\gamma{=}1$ is bracketed on either side by distributions that use the higher and lower thresholds.  The $\gamma{=}0$ distribution has a median date of 2044 and a 95 percent range from 2039 to 2053. The $\gamma{=}2$ distribution has a median date of 2033 and an earlier and slightly narrower 95 percent range from 20230 to 2039. The climate modeling literature has pointed to several factors that underpin the uncertainty in the timing of an initial September ice-free Arctic including natural climate variability, emissions path uncertainty, and uncertain sea ice dynamics \citep{SerrezeAndMeier2019}. These factors are at least partially accounted for in our analysis.

The multi-model mean CMIP5 climate projections described above are well outside the above-described distributional ranges. For a threshold of 1 million km\textsuperscript{2}, the mean projections reach a nearly ice-free Arctic in 2068 and 2089 under the RCP8.5 and RCP6.0 scenarios, respectively. The former date is denoted by a vertical brown bar in Figure \ref{densityS}. The range of dates across individual models is also extremely wide, stretching well over a century \citep{Jahn2016}. To narrow this range, researchers have omitted models with poor performance using a variety of sea ice metrics.  With such model selection procedures, the range of dates for a first nearly ice-free September is narrowed greatly to a 20-year span that runs from the 2040s to 2060s (\citealp{Massonnet2012}; \citealp{ThackHall2019}).\footnote{Again, the caveats of  \cite{Rosenblum2017} regarding such model selection are relevant.} Even such a carefully  circumscribed span is roughly a decade later than the simplified quadratic trend model produces. Moreover, the climate model simulation exercises are not designed to yield formal error estimates or measures of uncertainty as the spread of climate model forecasts in an ensemble is insufficient to completely characterize forecast uncertainty.

Finally, we note the availability of density forecasts for a variety of richer joint scenarios of interest. As one example, \cite{Jahn2016} describe other definitions of ``ice-free" that involve, for example, 5-year running means. Alternatively, ``ice-free" may require no ice for several consecutive months, for example, to accommodate meaningful freight shipping, tourism, mining, and commercial fishing \citep{Aksenov2017}. For example, the strong Autumn demand for international freight shipping to satisfy year-end Western holiday consumer demand could make a multi-month ice-free Arctic shipping lane of interest. In this case, the probability distribution of the initial occurrence of an ``ice-free" summer -- a joint ice-free August, September, and October -- could be relevant. Figure \ref{densityS} shows this distribution in black assuming an ice-free threshold of $\gamma{=}1$. This density is notably shifted right -- and more right-skewed -- compared to the September scenarios.

\section{Concluding Remarks}  \label{conclsec}

A rapidly warming Arctic is an ominous sign of the broader climate change caused by human activity, but declining Arctic sea ice also has an important influence on the pace and intensity of future climate change. Using statistical models, we have provided probabilistic projections of 21st-century Arctic sea ice that account for both intrinsic uncertainty and parameter estimation uncertainty. These projections indicate that summertime Arctic sea ice will quickly diminish and disappear -- with about a 60 percent probability of an effectively ice-free September Arctic occurring within two decades.

By contrast, the average projection from leading climate models implies an initial seasonal ice-free Arctic several decades later -- even assuming a business-as-usual emissions path. The slow and decreasing pace at which large-scale climate models reach an ice-free Arctic may be a serious shortcoming, and such conservative projections of sea ice loss could be a misleading guide for global climate policy.\footnote{Our statistical results are in line with the concerns of \cite{Stroeve2007} regarding the possibility that the slow projected decline in Arctic sea ice by climate models suggests they are underestimating the effects of greenhouse gases.  More recently, \cite{Guarinoetal2020} also emphasize deficiencies in climate models' Arctic representations.}

There are numerous examples, across many disciplines, showing that parsimonious statistical representations can provide forecasts that are at least as accurate as the ones from detailed structural models.\footnote{A classic example from economics is \cite{Nelson1972}, who showed that simple statistical models forecast the economy as well as large-scale structural models based on economic theory.} However, rather than treat statistical models as just forecast competitors to structural climate models, there is likely to be scope to use them as complementary representations going forward. The mechanisms governing Arctic sea ice loss and connecting that loss to atmospheric, oceanic, and permafrost responses are not fully captured in climate models. Statistical models may be able to help assist in bridging such gaps until a more complete understanding is available \citep{Parkinson2010}. 

In addition, statistical models may also be able to provide a benchmark for model performance that can be used for climate model evaluation, estimation/calibration, and selection. In particular,  our statistical approach summarizes a key aspect of the data, namely its historical dynamics and the associated forecast path (call it $PATH_D$, where $D$ stands for ``Data"). A  climate model with calibrated parameters $\theta$ produces a corresponding summary for the model  (call it $PATH_M(\theta)$, where $M$ stands for ``Model").  So, at least conceptually, we can consider
\begin{enumerate}
\item  \textit{evaluating} a climate model by examining a measure of divergence,  $|| PATH_D {-} PATH_M(\theta) ||$
\item  \textit{estimating} a climate model by finding  $\hat{\theta} =  argmin_{\theta} \, || PATH_D {-} PATH_M(\theta) ||$ (this is precisely indirect inference \citep{Gour1993}, using $PATH$ as the estimation window)
\item  \textit{selecting} a climate model by finding the model with smallest $|| PATH_D {-} PATH_M(\hat{\theta}) ||$, adjusted for d.f.  
\end{enumerate}
In practice such formal procedures are infeasible for massive global climate models, which attempt to address all aspects of the  high-dimensional climate state, of which sea ice is but one small part.  Hence, for example,  the ``right" window for complete  climate model estimation by indirect inference is much more complicated than just a sea ice path.  Less formal procedures directly motivated by the above considerations can nevertheless be highly revealing, as in our key Figure 6.

\clearpage

\appendix
\appendixpage
\addappheadtotoc
\newcounter{saveeqn}
\setcounter{saveeqn}{\value{section}}
\renewcommand{\theequation}{\mbox{\Alph{saveeqn}.\arabic{equation}}} \setcounter{saveeqn}{1}
\setcounter{equation}{0}

\section{Detailed Regression Estimation Results} \label{App1}

\renewcommand{\thefigure}{A\arabic{figure}}
\setcounter{figure}{0}

\renewcommand{\thetable}{A\arabic{table}}
\setcounter{table}{0}

\floatpagestyle{empty}
\begin{table} [h]
	\caption{Regression Estimation Results,  Quadratic Model with Restrictions}
	\begin{center}
		$
		SIE^*_t =   \sum_{i=1}^{12} \delta_i D_{it} + \sum_{j=1}^{12} \gamma_j D_{jt} {\cdot} TIME_t + \sum_{k=1}^{12} \alpha_k D_{kt} {\cdot} TIME_t^2  + \varepsilon_t
		$
		
		$\varepsilon_t =  \rho \varepsilon_{t-1} + v_t$
		
		$v_t \sim iid (0, \sigma^2)$
		
		$SIE_t = \max(SIE^*_t, 0)$
	\end{center}
	\label{vvv1}
	\begin{center}
		{
			\begin{tabular}{ c  c c  c c  c  c }
				\toprule
				&(1)&(2)&(3)&(4)&(5)&(6) \\
				&NONE & Seq &  NSeq &  Seq+NSeq & ALLeq & ALL0   \\
				\bottomrule
				$\delta_{1}$ &  15.1121* & 15.1372*  &   15.0865*  &    15.0922*  & 15.0373* & 15.2274*   \\
				$\delta_{2}$ & 15.9436* & 15.9570*  &  15.9404*  &    15.9457*  & 15.8894* & 16.0804*  \\
				$\delta_{3}$ &  16.0792* & 16.0820*  &   16.0049*  &    16.0100*  & 15.9524* & 16.1446*  \\
				$\delta_{4}$ &  15.3521* & 15.3443*  &   15.2278*  &    15.2326*  & 15.1741* & 15.3674*   \\
				$\delta_{5}$ &  13.8027* & 13.7832*  &   13.7604*  &    13.7650*  & 13.7056* & 13.9001*   \\
				$\delta_{6}$ &  12.4776* & 12.4443*  &   12.4352*  &    12.4397*  & 12.3796* & 12.5749*   \\
				$\delta_{7}$ &  10.4243* & 10.3733*  &   10.4804*  &    10.4849*  & 10.4243* & 10.6202*  \\
				$\delta_{8}$ &  8.0945* & 8.0202*  &   8.1441*  &    8.0754*  & 8.2547* & 8.4508*   \\
				$\delta_{9}$ &  7.4698* & 7.3645*  &   7.5179*  &    7.4201*  & 7.6022* & 7.7976*   \\
				$\delta_{10}$ &  9.1633* & 9.2520*  &   9.2146*  &   9.3080*  & 9.4932* & 9.6870*  \\
				$\delta_{11}$ &  11.4307* & 11.4872*  &   11.4928*  &    11.4997*  & 11.4488* & 11.6374*  \\
				$\delta_{12}$ &  13.5642* & 13.6042*  &   13.5673*  &    13.5735*  & 13.5205* & 13.7097*  \\
				\midrule
			\end{tabular}
		}
	\end{center}

\end{table}

\clearpage

\begin{table} [h]
	\caption*{Estimation Results,\ Quadratic Model with Restrictions (Continued)}
	\begin{center}
		{
			\begin{tabular}{ c  c c  c c  c  c }
				\toprule
				&(1)&(2)&(3)&(4)&(5)&(6) \\
				&NONE & Seq &  NSeq &  Seq+NSeq & ALLeq & ALL0   \\
				\bottomrule
				$\gamma_{1}$ &  -0.0026 & -0.0029*   &   -0.0023  & -0.0024*     & -0.0017 & -0.0040*   \\
				$\gamma_{2}$ &  -0.0023 & -0.0024  &   -0.0022*  & -0.0023*   & -0.0016 & -0.0039* \\
				$\gamma_{3}$ &  -0.0027 & -0.0028  &   -0.0018*  & -0.0019   & -0.0012 & -0.0035*   \\
				$\gamma_{4}$ &  -0.0031 & -0.0030  &   -0.0016  &  -0.0016   & -0.0009 & -0.0033*  \\
				$\gamma_{5}$ &  -0.0019 & -0.0016  &   -0.0014  &  -0.0014   & -0.0007 & -0.0031*  \\
				$\gamma_{6}$ &  -0.0028 & -0.0024  &   -0.0023*  &   -0.0024*  & -0.0016 & -0.0040*  \\
				$\gamma_{7}$ &  -0.0034* & -0.0028*  &   -0.0041*  &  -0.0042* & -0.0034* & -0.0058*   \\
				$\gamma_{8}$ &  -0.0021 & -0.0012  &   -0.0027  &   -0.0019  & -0.0040* & -0.0064*  \\
				$\gamma_{9}$ &  -0.0030* & -0.0018  &  -0.0036*  &  -0.0024* & -0.0045* & -0.0069*  \\
				$\gamma_{10}$ &  -0.0006 & -0.0016  &   -0.0012  &   -0.0023*  & -0.0044* & -0.0068* \\
				$\gamma_{11}$ &  -0.0021 & -0.0028*  &   -0.0029*  &  -0.0030*  & -0.0023* & -0.0046*  \\
				$\gamma_{12}$ &  -0.0023 & -0.0028  &   -0.0024*  &    -0.0024* & -0.0017 & -0.0041*  \\
				\midrule
				$\alpha_{1}$ &  -2.72E-06 & -2.11E-06  & \textbf{-3.40E-06}   &    \textbf{-3.29E-06}  & \textbf{-4.78E-06}* & \textbf{0}  \\
				$\alpha_{2}$ &  -3.31E-06 & -2.99E-06  &  \textbf{-3.40E-06}   &   \textbf{-3.29E-06}  & \textbf{-4.78E-06}* & \textbf{0}  \\
				$\alpha_{3}$ &  -1.53E-06 & -1.47E-06  &   \textbf{-3.40E-06}  &   \textbf{-3.29E-06}  & \textbf{-4.78E-06}* & \textbf{0}  \\
				$\alpha_{4}$ &  -3.24E-07 & -5.17E-06  &   \textbf{-3.40E-06}  &   \textbf{-3.29E-06}  & \textbf{-4.78E-06}* & \textbf{0}   \\
				$\alpha_{5}$ &  -2.38E-06 & -2.84E-06  &   \textbf{-3.40E-06}  &   \textbf{-3.29E-06}  &\textbf{-4.78E-06}* & \textbf{0}  \\
				$\alpha_{6}$ &  -2.39E-06 & -3.18E-06  &   \textbf{-3.40E-06}  &   \textbf{-3.29E-06}  & \textbf{-4.78E-06}* & \textbf{0}  \\
				$\alpha_{7}$ &  -4.78E-06 & -5.96E-06*  &   \textbf{-3.40E-06}  &    \textbf{-3.29E-06}  & \textbf{-4.78E-06}* & \textbf{0}  \\
				$\alpha_{8}$ &  -8.57E-06* & \textbf{-1.03E-05*}  &   -7.35E-06*  &  \textbf{-8.96E-06}*  & \textbf{-4.78E-06}* & \textbf{0}  \\
				$\alpha_{9}$ &  -7.89E-06* & \textbf{-1.03E-05*}  &   -6.69E-06*  &    \textbf{-8.96E-06}* & \textbf{-4.78E-06}* & \textbf{0}  \\
				$\alpha_{10}$ &  -1.24E-05* & \textbf{-1.03E-05*}  &   -1.11E-05*  &    \textbf{-8.96E-06}*  & \textbf{-4.78E-06}* & \textbf{0} \\
				$\alpha_{11}$ &  -4.99E-06*  &   -3.57E-06  &   \textbf{-3.40E-06}  & \textbf{-3.29E-06} & \textbf{-4.78E-06}* & \textbf{0} \\
				$\alpha_{12}$ &  -3.45E-06 & -2.48E-06  &   \textbf{-3.40E-06}  &    \textbf{-3.29E-06}  & \textbf{-4.78E-06}* & \textbf{0} \\
				\midrule
				$\rho$ &  0.7302* & 0.7284*  &   0.7298*  &    0.7270*  & 0.7288* & 0.7461*  \\
				$\sigma^2$  &  0.0468* & 0.0473*  &   0.0472*  &  0.0478*  & 0.0491* & 0.0497*   \\
				\midrule
				$R^2$ & 0.996 &0.995 & 0.995& 0.995 & 0.995& 0.995\\
				$DW$ & 1.71 & 1.71 &1.71 & 1.71 &1.71 & 1.72\\
				$Skew$ & -0.17  &-0.19  & -0.16 & -0.20  &-0.18  & -0.14\\
				$Kurt$ & 3.77  & 3.75  & 3.70  & 3.70 & 3.93  & 3.95\\
				\bottomrule	
			\end{tabular}
		}
	\end{center}
		\end{table}

\clearpage
\begin{center}
	Estimation Results, Quadratic Model with Restrictions (Continued)
\end{center}

\bigskip

	\begin{spacing}{1.0}  \noindent Notes to table:  We show maximum-likelihood  estimates of the  parameters of the general quadratic model (\ref{eq:Shadow2}), with various equality constraints imposed on the quadratic coefficients $\alpha_1, ..., \alpha_{12}$. ``NONE" denotes no constraints, which corresponds to the general quadratic model (\ref{eq:Shadow2}).  ``Seq" (``Summer equal") denotes August-October equal  ($\alpha_{8} {=} \alpha_{9}  {=}  \alpha_{10}$).  ``NSeq" (``Non-Summer equal") denotes November-July equal ($\alpha_{11}  {=}   \alpha_{12}  {=}  \alpha_{1}  {=}  ...  {=}  \alpha_{7}$). ``Seq+NSeq" denotes summer months equal and non-summer months (separately) equal, which corresponds to the simplified quadratic model.  ``ALLeq" denotes all months equal ($\alpha_{1}  {=}  ...  {=}  \alpha_{12}$). ``ALL0" denotes all months 0 ($\alpha_{1}  {=}  ...  {=}  \alpha_{12} {=} 0$),  which corresponds to the linear model (\ref{eq:SIE}).  In each column, we show constrained parameter estimates in boldface. ``*" denotes significance at the ten percent level.
\end{spacing}

\bigskip

\begin{figure}[h!]
	\caption{$SIE$, Actual and Residual}
	\begin{center}
		\includegraphics[trim= 6mm 50mm 0mm 50mm, clip, scale=.08]{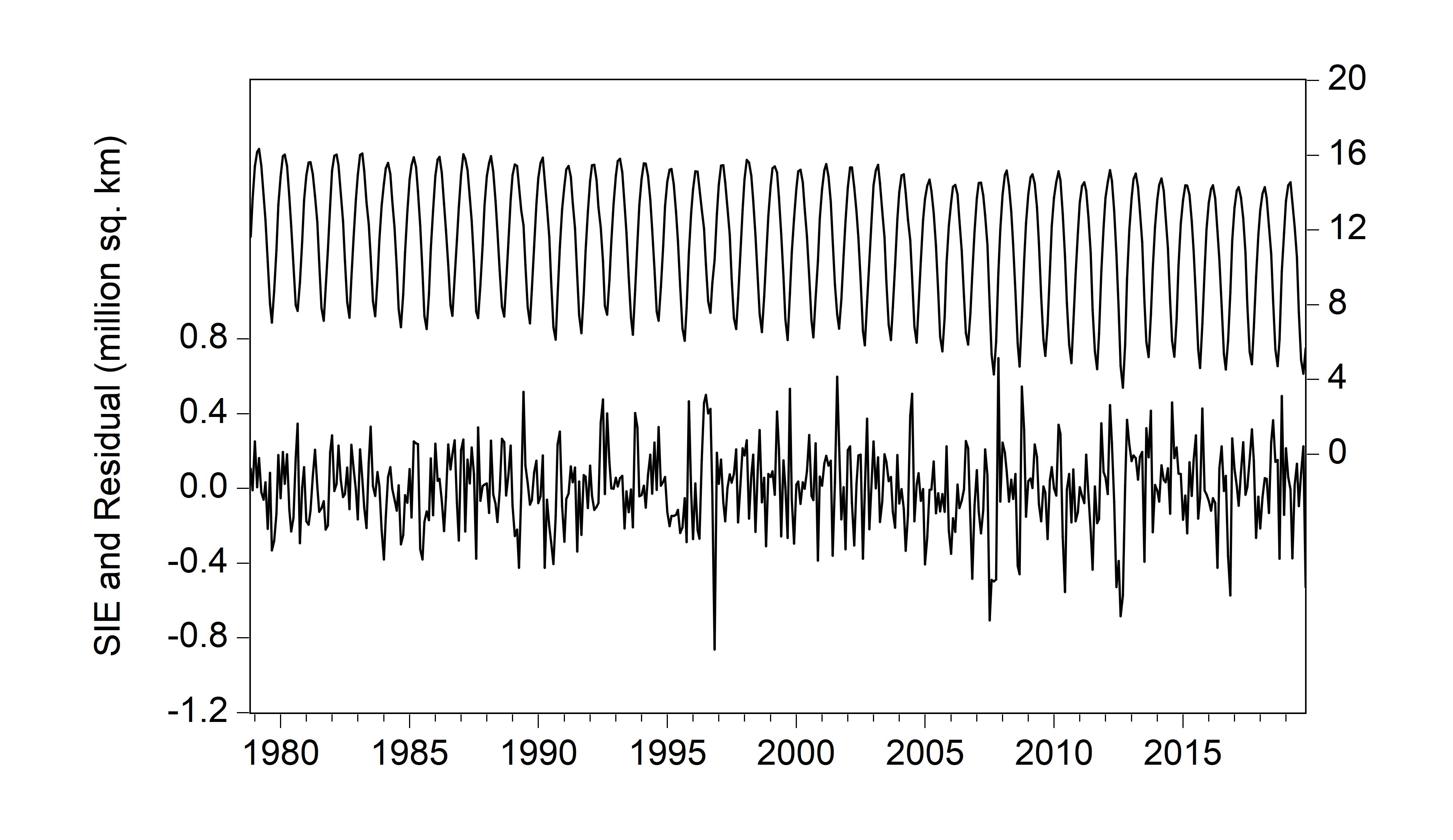}
		\includegraphics[trim= 6mm 50mm 0mm 50mm, clip, scale=.08]{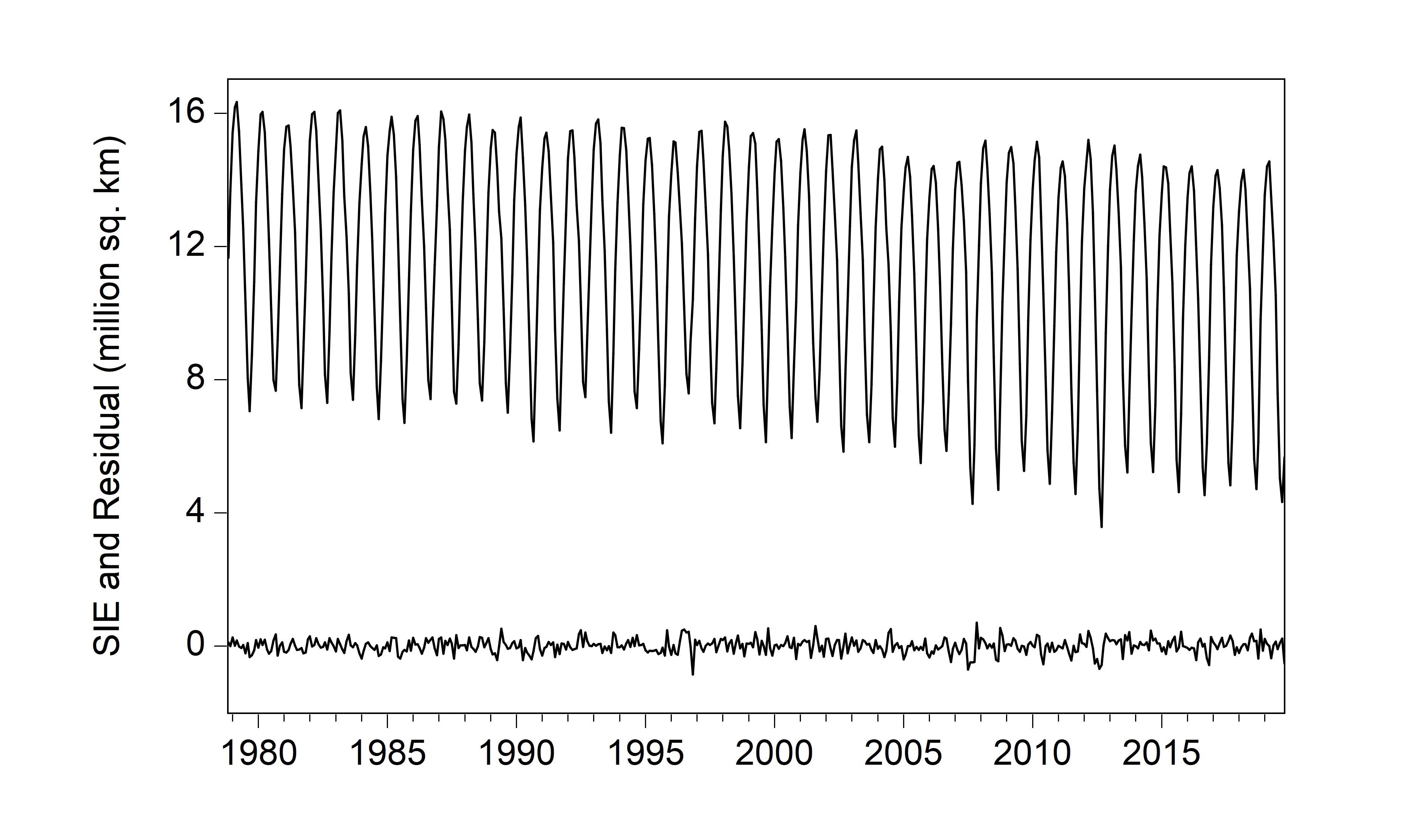}
		\label{tsplot22}
	\end{center}
	\begin{spacing}{1.0} \footnotesize \noindent Notes: The left panel uses different scales (actual on the right, residual on the left), so the residual is greatly enlarged.  The right panel uses a common scale, making clear that the residual is negligible.
	\end{spacing}
\end{figure}

\begin{figure}[h!]
	\caption{$SIE$ Residual Histogram and Best-Fitting Gaussian}
	\begin{center}
		\includegraphics[trim= 6mm 50mm 0mm 50mm, clip, scale=.08]{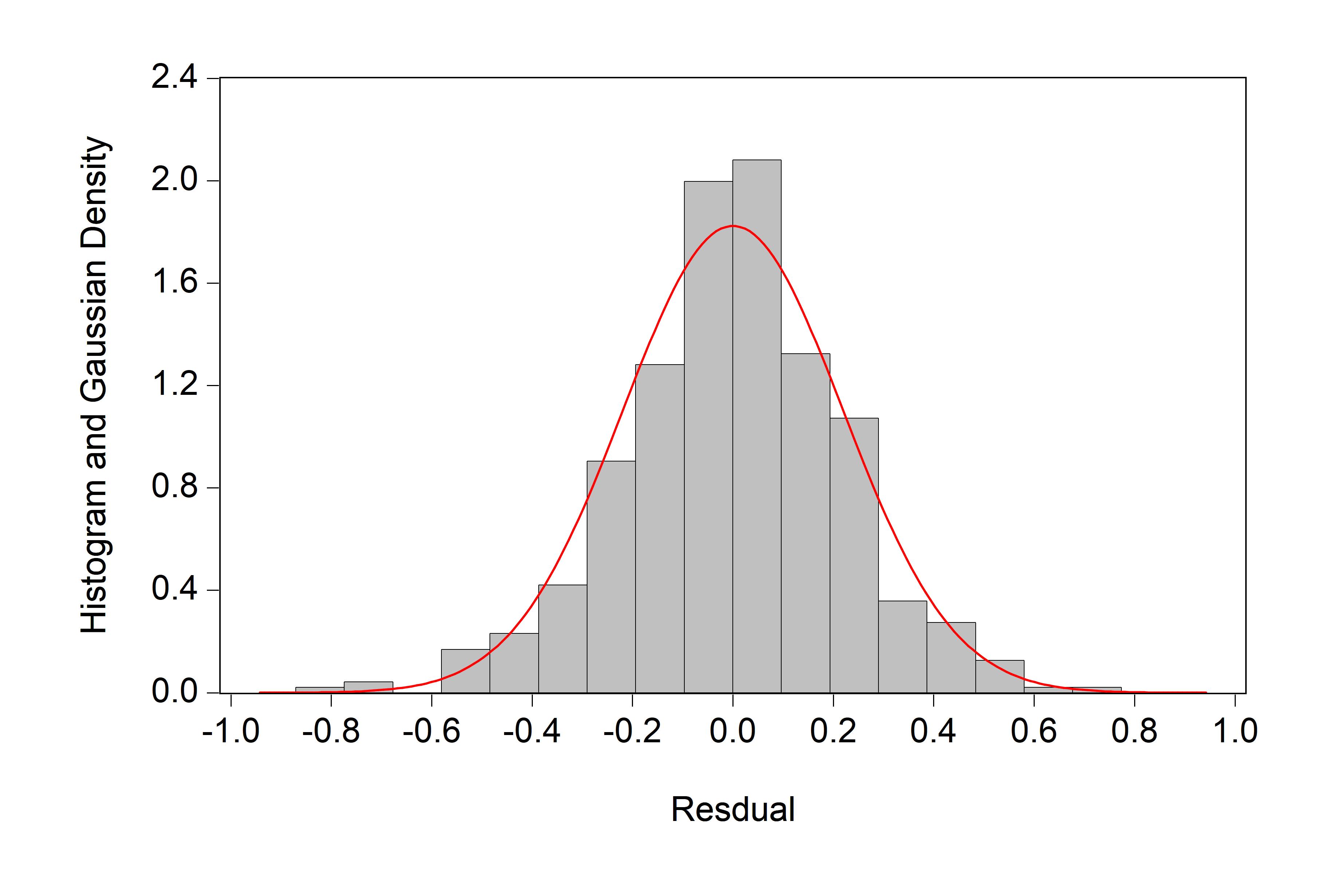}
		\label{tsplot32}
	\end{center}
	\begin{spacing}{1.0} \footnotesize \noindent Notes: We show the  residual histogram, together with a Gaussian density fit to the residuals.
	\end{spacing}
\end{figure}

\clearpage

\section{Simulation Methods} \label{App2}

We  estimate scenario probability distributions using the simplified quadratic model (that is, the quadratic model (\ref{eq:Shadow2}) subject to the constraints $\alpha_{8} {=} \alpha_{9}  {=}  \alpha_{10}$ and $\alpha_{11}  {=}   \alpha_{12}  {=}  \alpha_{1}  {=}  ...  {=}  \alpha_{7}$) and a  simulation procedure that accounts for parameter estimation uncertainty and allows for   potentially non-Gaussian  serially correlated stochastic shocks. To describe this simulation procedure, it is useful to re-write the quadratic model (\ref{eq:Shadow2}) in a more concise notation:
$$
SIE^*_t = x'_t  \beta + \varepsilon_t
$$
$$
\varepsilon_t =  \rho \varepsilon_{t-1} + v_t
$$
$$
v_t \sim iid (0, \sigma^2)
$$
$$
SIE_t = \max(SIE^*_t, 0),
$$
where $x'_t {=}(D_{1t}, ..., D_{12t},~
D_{1t} {\cdot} TIME_t, ..., D_{12t} {\cdot} TIME_t,~
D_{1t} {\cdot} TIME_t^2, ..., D_{1t} {\cdot} TIME_t^2)$ and

\noindent $\beta' {=}(\delta_1,..., \delta_{12}, \gamma_1, ..., \gamma_{12}, \alpha_1, ..., \alpha_{12})$. We estimate the model by  maximum likelihood using the historical data sample $t = 1, ..., T$, as discussed earlier, and then we  simulate $i=1, ..., 10,000$ future paths based on the estimated model.  Simulation $i$ proceeds as follows:

\begin{enumerate}
	
	\item Draw  ${\beta}^{(i)}$ from the sampling distribution of the estimated parameter vector $\hat{\beta}$, and form $x'_t {\beta}^{(i)}$, $t = T{+}1, ..., 2099M12$.
	
	\item  Draw $v_t^{(i)}$ by sampling with replacement from the observed  $\hat{v}_t$'s (with equal weights $1/T$), $t = T{+}1, ..., 2099M12$.
	
	\item  Draw  ${\rho}^{(i)}$ from the sampling distribution of  $\hat{\rho}$, and form $\varepsilon_t^{(i)} = {\rho}^{(i)} \varepsilon _{t-1} + v_t^{(i)}$, $t = T{+}1, ..., 2099M12$, using the last observed historical residual $\hat{\varepsilon}_T$ as the initial condition.
	
	\item Form $SIE_t^{(i)} = x'_t {\beta}^{(i)} + {\varepsilon}^{(i)}_t$, $t = T{+}1, ..., 2099M12$.
	
\end{enumerate}

\noindent Finally, we estimate event probabilities of interest as the relative frequency of occurrence across the simulated paths.

\clearpage
\bibliographystyle{Diebold}
\addcontentsline{toc}{section}{References}
\bibliography{Bibliography}

\end{document}